\documentclass[12pt]{article}

\author{Yu.~M.~Zinoviev
       \thanks{E-mail address: yurii.zinoviev@ihep.ru} \\
               {\it Institute for High Energy Physics} \\
               {\it Protvino, Moscow Region, 142280,Russia}}
\title{On Dual Formulation of Gravity}

\date{}
 
\begin{document}

\maketitle

\begin{abstract}
In this paper we consider a possibility to construct dual formulation
of gravity where the main dynamical field is the Lorentz connection
$\omega_\mu{}^{ab}$ and not that of tetrad $e_\mu{}^a$ or metric
$g_{\mu\nu}$. Our approach is based on the usual dualization
procedure which uses first order parent Lagrangians but in (Anti) de
Sitter space and not in the flat Minkowski one. It turns out that
in $d=3$ dimensions such dual formulation is related with the so
called exotic parity-violating interactions for massless spin-2
particles.
\end{abstract}

\thispagestyle{empty}
\newpage
\setcounter{page}{1}

\section*{Introduction}

Investigations of dual formulations for tensor fields are important
for understanding of alternative formulations of known theories like
gravity as well as understanding of their role in superstrings.
Common procedure for obtaining such dual formulations is based on the
parent first order Lagrangians. As is well known in flat Minkowski
space such dualization procedure leads to different results for
massive and massless particles. At the same time in (Anti) de Sitter
space-time gauge invariance requires introduction quadratic mass-like
terms into the Lagrangians. As a result dualization for massless
particles in (Anti) de Sitter spaces \cite{MV04} goes exactly in the
same way as that for massive particles \cite{Zin05} and gives
results different from ones for dualization of massless particles
\cite{BCH03} in flat Minkowski space.

In this paper using such dualization procedure we consider a
possibility to construct dual formulation of gravity where the main
dynamical quantity is a Lorentz connection field $\omega_\mu{}^{ab}$.
Note that at the Hamiltonian level such description was discussed in
\cite{CZ99}. Also such dual formulation of $d=3$ gravity was recently
discussed in \cite{Jul05}. It turns out that in $d=3$ dimensions
such dual formulation of gravity is related with the so called exotic
parity-violating interactions for massless spin-2 particles
\cite{BG00,Anco03}. So we start with $d=3$ case and show that such
exotic interaction can be viewed as higher derivatives interactions
in terms Lorentz connection $\omega_\mu{}^{ab}$. Then we show how
such interaction could be obtained from the usual gravitational
interactions by dualization procedure starting with (Anti) de Sitter
space and then considering a kind of flat limit. Then in the next
section we consider straightforward generalization of such theory on
arbitrary $d \ge 4$ dimensions.

\section{Dual gravity in $d=3$}

Investigations of possible interactions for massless spin-2 particles
have shown that in $d=3$ case there exist non-trivial "exotic
parity-violating" higher derivatives interactions \cite{BG00,Anco03}.
The simplest way to see this \cite{Zin03} is to start with the first
order formulation for massless spin-2 particle using "triad"
$h_\mu{}^a$ and Lorentz connection $\omega_\mu{}^{ab}$ and introduce
dual variable $f_\mu{}^a = \frac{1}{2} \varepsilon^{abc}
\omega_\mu{}^{bc}$. In this notations the Lagrangian for free
massless spin-2 particles has a very simple form:
\begin{equation}
{\cal L}_0 = \frac{1}{2} \left\{ \phantom{|}^{\mu\nu}_{ab} \right\}
f_\mu{}^a f_\nu{}^b - \varepsilon^{\mu\nu\alpha} f_\mu{}^a
\partial_\nu h_\alpha{}^a 
\end{equation}
Here
$$
\left\{ \phantom{|}^{\mu\nu}_{ab} \right\} = \delta_a{}^\mu
\delta_b{}^\nu - \delta_a{}^\nu \delta_b{}^\mu
$$
and so on. This Lagrangian is invariant under the following local
gauge transformations:
\begin{equation}
\delta h_{\mu a} = \partial_\mu \xi_a + \varepsilon_{\mu a b} \eta^b
\qquad \delta f_\mu{}^a = \partial_\mu \eta_a 
\end{equation}
Then it is easy to check that if we add the following cubic terms to
the Lagrangian and appropriate corrections to gauge transformation
laws:
\begin{equation}
{\cal L}_1 = - \frac{K}{6}  \left\{ \phantom{|}^{\mu\nu\alpha}_{abc}
\right\} f_\mu{}^a f_\nu{}^b f_\alpha{}^c \qquad \delta_1 h_\mu{}^a =
- K \varepsilon^{abc} f_\mu{}^b \eta^c
\end{equation}
where $K$ --- arbitrary coupling constant, we obtain gauge invariant
interacting theory. In this, equation of motion for the $f_\mu{}^a$
field are still algebraic, but non-linear now. So if we try to solve
this equation in passing to second order formulation we get
essentially non-linear theory with higher and higher derivatives
terms. To see what kind of theory we get let us consider lowest order
approximations. It will be convenient to introduce "dual torsion"
$$
 T^{\mu a} = - \varepsilon^{\mu\nu\alpha} \partial_\nu h_\alpha{}^a,
\qquad \hat{T}^{\mu a} = T^{\mu a} - e^{\mu a} T 
$$
Then from the quadratic Lagrangian we easily obtain:
$$
f_\mu^{(1)a} = \hat{T}^a{}_\mu, \qquad {\cal L}_0 = - \frac{1}{2}
\left\{ \phantom{|}^{\mu\nu}_{ab} \right\} \hat{T}^a{}_\mu
\hat{T}^b{}_\nu
$$
In the next quadratic order we get:
$$
f_\mu^{(2)a} = - K \hat{T}^a{}_b \hat{T}^b{}_\mu + \frac{K}{4}
e_\mu{}^a [\hat{T}^b{}_c \hat{T}^c{}_b + \hat{T}^2]
$$
Substituting this expressions back to the first order Lagrangian and
keeping only terms cubic in fields we obtain an interactions in a
first non-trivial order:
\begin{equation}
{\cal L}_1 = \frac{K}{6} \left\{ \phantom{|}^{\mu\nu\alpha}_{abc}
\right\} \hat{T}^a{}_\mu \hat{T}^b{}_\nu \hat{T}^c{}_\alpha
\end{equation}
and this is just the interaction considered in \cite{BG00,Anco03}.
Note here that such interactions do not necessarily violate parity
because one can always assign $f_\mu{}^a$ to be a tensor, while
$h_\mu{}^a$ --- a pseudotensor. Now we can once again use a
peculiarity of $d=3$ space and dualize $h_\mu{}^a$ instead of
$f_\mu{}^a$:
$h_\alpha{}^a = \frac{1}{2} \varepsilon^{abc} \omega_\alpha{}^{bc}$
Then we can rewrite all results in terms of this new variable by
noting that:
$$
\hat{T}^a{}_\mu = (R_\mu{}^a - \frac{1}{4} \delta_\mu{}^a R) = 
\hat{R}_\mu{}^a
$$
where we have introduced usual field strength:
$$
R_{\mu\nu}{}^{ab} = \partial_\mu \omega_\nu{}^{ab} - \partial_\nu
\omega_\mu{}^{ab}, \qquad R_\mu{}^a = \delta^\nu{}_b R_{\mu\nu}{}^{ab}
$$
Now the cubic interactions looks like:
\begin{equation}
{\cal L}_1 =  \frac{K}{6} \left\{ \phantom{|}^{\mu\nu\alpha}_{abc}
\right\} \hat{R}_\mu{}^a \hat{R}_\nu{}^b \hat{R}_\alpha{}^c
\end{equation}

In $d=3$ dimensions all these look just like trivial field
redefinition, but looking this way it has to be clear that there
should exist a generalization of such interactions on arbitrary 
$d \ge 4$. To see how this generalization could be constructed we
have to reobtain the same results without use of peculiarities of
$d=3$ dimensions. Now we will show that it is indeed possible by
following usual dualization procedure based on the parent first order
Lagrangians. Crucial fact here is that dualization for massless
particles in (Anti) de Sitter spaces \cite{MV04} goes in way similar
to the one for massive particles in flat space \cite{Zin05} and not
to that for massless ones \cite{BCH03}. So let us return back to the
free case and start with massless particle in (Anti) de Sitter
background space. A first order Lagrangians looks now as follows:
\begin{equation}
{\cal L}_0 = \frac{1}{2} \left\{ \phantom{|}^{\mu\nu}_{ab} \right\}
f_\mu{}^a f_\nu{}^b - \varepsilon^{\mu\nu\alpha} f_\mu{}^a D_\nu
h_\alpha{}^a + \frac{\kappa}{2} 
\left\{ \phantom{|}^{\mu\nu}_{ab} \right\} h_\mu{}^a h_\nu{}^b
\end{equation}
and is invariant under the following local gauge transformations:
\begin{equation}
\delta h_{\mu a} = D_\mu \xi_a + \varepsilon_{\mu a b} \eta^b \qquad
\delta f_\mu{}^a = D_\mu \eta_a + \kappa \varepsilon_{\mu a b} \xi^b
\end{equation}

Working with the first order formalism it is very convenient to 
use tetrad formulation of the underlying (Anti) de Sitter space.
We denote tetrad as $e_\mu{}^a$ (let us stress that it is not a
dynamical quantity here, just a background field) and Lorentz
covariant derivative as $D_\mu$. (Anti) de Sitter space is a 
constant curvature space with zero torsion, so we have:
\begin{equation}
D_{[\mu} e_{\nu]}{}^a = 0, \qquad [ D_\mu, D_\nu ] v^a =
\kappa (e_\mu{}^a e_\nu{}^b - e_\mu{}^b e_\nu{}^a) v_b
\end{equation} 
where $\kappa = - 2 \Lambda/(d-1)(d-2)$.

Now we switch on usual gravitational interaction by adding to the
Lagrangian the following cubic terms:
\begin{equation}
{\cal L}_1 = - \frac{k}{2}  \left\{ \phantom{|}^{\mu\nu\alpha}_{abc}
\right\} f_\mu{}^a f_\nu{}^b h_\alpha{}^c - \frac{k \kappa}{6}
 \left\{ \phantom{|}^{\mu\nu\alpha}_{abc} \right\} h_\mu{}^a
h_\nu{}^b h_\alpha{}^c
\end{equation}
as well as appropriate corrections to gauge transformation laws:
\begin{equation}
\delta_1 h_\mu{}^a = k \varepsilon^{abc} ( f_\mu{}^b \xi^c +
 h_\mu{}^b \eta^c) \qquad
\delta_1 f_\mu{}^a = k \varepsilon^{abc} ( f_\mu{}^b \eta^c + 
\kappa h_\mu{}^b \xi^c)
\end{equation}
Note that in $d=3$ case this gives us complete interacting theory. 
Then we switch back to the usual variable: $f_\mu{}^a = \frac{1}{2}
\varepsilon^{abc} \omega_\mu{}^{bc}$. Also in order to have canonical
normalization of fields in dual theory (where $\omega$ is main
dynamical quantity now, while $h$ --- just auxiliary field) we make a
rescaling : $\omega \rightarrow \sqrt{\kappa} \omega$ and
$h \rightarrow \frac{1}{\sqrt{\kappa}} h$. In this a quadratic
Lagrangian takes the form:
\begin{equation}
{\cal L}_0 = \frac{\kappa}{2} \left\{ \phantom{|}^{\mu\nu}_{ab}
\right\} \omega_\mu{}^{ac} \omega_\nu{}^{bc} - \frac{1}{2}
 \left\{ \phantom{|}^{\mu\nu\alpha}_{abc} \right\}
 \omega_\mu{}^{ab} D_\nu h_\alpha{}^c  + \frac{1}{2}
 \left\{ \phantom{|}^{\mu\nu}_{ab} \right\} h_\mu{}^a h_\nu{}^b
\end{equation}
and gauge transformations leaving it invariant (now $\eta^a =
\frac{1}{2} \varepsilon^{abc} \eta_{bc}$)
\begin{equation}
\delta h_{\mu a} = D_\mu \xi_a + \kappa \eta_{\mu a} \qquad
\delta \omega_\mu{}^{ab} = D_\mu \eta^{ab} - e_\mu{}^a \xi^b +
e_\mu{}^b \xi^a
\end{equation}
At the same time an interacting Lagrangian in these variables looks
like:
\begin{equation}
{\cal L}_1 = - \frac{k\sqrt{\kappa}}{2} \left\{ 
\phantom{|}^{\mu\nu\alpha}_{abc} \right\} \omega_\mu{}^{ad}
\omega_\nu{}^{bd} h_\alpha{}^c - \frac{k}{6\sqrt{\kappa}}
 \left\{ \phantom{|}^{\mu\nu\alpha}_{abc} \right\} h_\mu{}^a
 h_\nu{}^b h_\alpha{}^c
\end{equation}
with appropriate corrections for gauge transformations:
\begin{eqnarray}
\delta_1 h_\mu{}^a &=& - k \sqrt{\kappa} ( \omega_\mu{}^{ab} \xi^b +
h_{\mu b} \eta^{ba}) \nonumber \\
\delta_1 \omega_\mu{}^{ab} &=& - k \sqrt{\kappa} ( \omega_\mu{}^{ac}
\eta^{cb} - \omega_\mu{}^{bc} \eta^{ca}) + \frac{k}{\sqrt{\kappa}}
(h_\mu{}^a \xi^b - h_\mu{}^b \xi^a)
\end{eqnarray}

Usually in passing to the second order formulation one solves
algebraic equation of motion for the $\omega$ field (which
geometrically give zero torsion condition). Then putting results back
into the initial first order Lagrangian one obtains ordinary second
order formulation in terms of (symmetric) tensor field. Here we
proceed another way and try to solve equation for $h$ field which is
also algebraic in (Anti) de Sitter background. This equation looks as:
\begin{equation}
\frac{\delta {\cal L}}{\delta h_\mu{}^a} = - \frac{1}{4}
\left\{ \phantom{|}^{\mu\nu\alpha}_{abc} \right\} R_{\nu\alpha}{}^{bc}
+ \left\{ \phantom{|}^{\mu\nu}_{ab} \right\}  h_\nu{}^b -
\frac{k}{2\sqrt{\kappa}} \left\{ \phantom{|}^{\mu\nu\alpha}_{abc}
\right\} h_\nu{}^b h_\alpha{}^c - \frac{k\sqrt{\kappa}}{2} 
\left\{ \phantom{|}^{\mu\nu\alpha}_{abc} \right\} \omega_\nu{}^{bd}
\omega_\alpha{}^{cd}
\end{equation}
where now $R_{\mu\nu}{}^{ab} = D_\mu \omega_\nu{}^{ab} - D_\nu
\omega_\mu{}^{ab}$. In the lowest order approximation we get:
\begin{equation}
h_\mu^{(1)a} = \hat{R}_\mu{}^a, \qquad \hat{R}_\mu{}^a = R_\mu{}^a -
\frac{1}{4} e_\mu{}^a R
\end{equation}
while a second order quadratic Lagrangian takes the form:
\begin{equation}
{\cal L}_0 = - \frac{1}{2} \left\{ \phantom{|}^{\mu\nu}_{ab} \right\}
\hat{R}_\mu{}^a \hat{R}_\nu{}^b + \frac{\kappa}{2}
\left\{ \phantom{|}^{\mu\nu}_{ab} \right\} \omega_\mu{}^{ac}
\omega_\nu{}^{bc}
\end{equation}
Note, that appearance of quadratic mass-like terms is natural in
(Anti) de Sitter background and does not mean that $\omega$ field
becomes massive. It is important that besides usual gauge
transformations $\delta \omega_\mu{}^{ab} = D_\mu \eta^{ab}$ this
Lagrangian also invariant under the local shifts
$\delta \omega_\mu{}^{ab} = - e_\mu{}^a \xi^b + e_\mu{}^b \xi^a$
which is a remnant of $\xi$-invariance of initial first order
Lagrangian. To check this invariance one can use that under these
transformations we have $\delta \hat{R}_\mu{}^a = D_\mu \xi^a$.

Now we proceed and consider next approximation with cubic interaction
terms in the Lagrangian and linear terms in gauge transformation
laws. Before we give explicit formulas let us discuss what kind of
theory we obtain. Schematically the solution of $h$ equation and cubic
Lagrangian look like:
\begin{eqnarray}
h^{(2)} &\sim& \frac{k}{\sqrt{\kappa}} (D \omega) (D \omega) +
k \sqrt{\kappa} \omega \omega \nonumber \\
{\cal L}_1 &\sim& \frac{k}{\sqrt{\kappa}} (D \omega) (D \omega)
(D \omega) + k \sqrt{\kappa} (D \omega) \omega \omega 
\end{eqnarray}
So the "main" interaction terms are cubic three derivatives ones
constructed from the gauge invariant field strengths $(D\omega)$, the
coupling constant being $K = \frac{k}{\sqrt{\kappa}}$ and at this
level theory is essentially abelian. Only the presence of nonzero
cosmological term adds one derivative Yang-Mills type coupling with
dimensionless coupling constant being $g = k\sqrt{\kappa}$. In this,
our theory becomes non-abelian, the gauge group being the Lorentz
group. The non-trivial interactions given above could be reproduced
now in a kind of "flat" limit when $k \rightarrow 0$ and $\kappa
\rightarrow 0$ keeping $K$ fixed. Indeed, in this limit we obtain:
\begin{equation}
h_\mu^{(2)a} = K \left[ - \hat{R}_\mu{}^b \hat{R}_b{}^a +
\hat{R}_\mu{}^a \hat{R} + \frac{1}{4} e_\mu{}^a \hat{R}_b{}^c
\hat{R}_c{}^b - \frac{1}{4} e_\mu{}^a \hat{R}^2 \right]
\end{equation}
while the cubic terms in the Lagrangian take the same simple form as
before:
\begin{equation}
{\cal L}_1 = - \frac{K}{6} \left\{ \phantom{|}^{\mu\nu\alpha}_{abc}
\right\} \hat{R}_\mu{}^a \hat{R}_\nu{}^b \hat{R}_\alpha{}^c
\end{equation}
Besides the trivial at these limit invariance under the $\eta^{ab}$
gauge transformations this Lagrangian is also invariant under the
local shifts $\xi^a$ with appropriate corrections:
$$\delta_1 \omega_\mu{}^{ab} = K(\hat{R}_\mu{}^a \xi^b - 
\hat{R}_\mu{}^b \xi^a)$$
Let us stress that it is the invariance under these shifts that fixes
the particular structure of cubic interactions among many other
possible gauge invariant terms that could be easily constructed.

\section{Dual gravity in $d \ge 4$}

In this section we consider straightforward generalization of the
procedure given above to the case of arbitrary $d \ge 4$ space-times.
Again we start with the first order formulation of massless spin-2
particle in (Anti) de Sitter background with the Lagrangian:
\begin{equation}
{\cal L}_0 = \frac{\kappa}{2} \left\{ \phantom{|}^{\mu\nu}_{ab}
\right\} \omega_\mu{}^{ac} \omega_\nu{}^{bc} - \frac{1}{2}
 \left\{ \phantom{|}^{\mu\nu\alpha}_{abc} \right\} D_\mu
\omega_\nu{}^{ab} h_\alpha{}^c  + \frac{d-2}{2} \left\{
\phantom{|}^{\mu\nu}_{ab} \right\}  h_\mu{}^a h_\nu{}^b
\end{equation}
Here we have already made a rescaling of fields appropriate for dual
version. Then we add the usual gravitational interactions at the first
non-trivial (cubic) order:
\begin{equation}
{\cal L}_1 = \frac{k\sqrt{\kappa}}{2} \left\{ 
\phantom{|}^{\mu\nu\alpha}_{abc} \right\} \omega_\mu{}^{ad}
\omega_\nu{}^{bd} h_\alpha{}^c - \frac{k}{4\sqrt{\kappa}}
\left\{ \phantom{|}^{\mu\nu\alpha\beta}_{abcd} \right\} D_\mu
\omega_\nu{}^{ab} h_\alpha{}^c h_\beta{}^d  +
\frac{(2d-5)k}{6\sqrt{\kappa}}  \left\{
\phantom{|}^{\mu\nu\alpha}_{abc} \right\} h_\mu{}^a h_\nu{}^b
h_\alpha{}^c
\end{equation}
As is well known working with tetrad formulation of gravity and
especially with supergravity theories it is very convenient to use
the so called "1 and 1/2" order formalism. But here to construct a
dual theory we have to work in a "honest" first order formalism
taking into account gauge transformations for all fields. In this
approximation they have the following form:
\begin{eqnarray}
\delta_1 \omega_\mu{}^{ab} &=& \frac{k}{\sqrt{\kappa}} [ \xi^\nu
R_{\nu\mu}{}^{ab} + (R_\mu{}^a \xi^b - R_\mu{}^b \xi^a) +
\frac{1}{d-2} \xi^\nu (e_\mu{}^a R_\nu{}^b - e_\mu{}^b R_\nu{}^a) -
\nonumber \\
 && - \frac{1}{2(d-2)} (e_\mu{}^a \xi^b - e_\mu{}^b \xi^a) R -
(d-2)(h_\mu{}^a \xi^b -   h_\mu{}^b \xi^a)] \\
\delta_1 h_\mu{}^a &=& k \sqrt{\kappa} \omega_\mu{}^{ab} \xi^b
\nonumber
\end{eqnarray}
for the $\xi^a$-transformations as well as
\begin{equation}
\delta_1 \omega_\mu{}^{ab} = k\sqrt{\kappa} (\omega_\mu{}^{ac} 
\eta^{cb} -  \omega_\mu{}^{bc} \eta^{ca}) \qquad
\delta_1 h_\mu{}^a = k\sqrt{\kappa} h_{\mu b} \eta^{ba}
\end{equation}
for the $\eta^{ab}$-ones. Note that the main difference from the
$d=3$ case is rather complicated form for the $\xi$-transformations of
$\omega$ field. As we will see this leads to the essential difference
in the structure of interacting Lagrangian. Now we try to solve
algebraic equation for $h$ field which in this approximation looks as
follows:
\begin{eqnarray}
\frac{\delta {\cal L}}{\delta h_\mu{}^a} &=& - \frac{1}{4}
\left\{ \phantom{|}^{\mu\nu\alpha}_{abc} \right\} R_{\nu\alpha}{}^{bc}
 + (d-2) \left\{ \phantom{|}^{\mu\nu}_{ab} \right\}  h_\nu{}^b -
\frac{k}{4\sqrt{\kappa}} \left\{
\phantom{|}^{\mu\nu\alpha\beta}_{abcd} \right\}
R_{\nu\alpha}{}^{bc} h_\beta{}^d + \nonumber \\
 && + \frac{(2d-5)k}{2\sqrt{\kappa}} \left\{
\phantom{|}^{\mu\nu\alpha}_{abc}  \right\} h_\nu{}^b h_\alpha{}^c +
\frac{k\sqrt{\kappa}}{2} \left\{ \phantom{|}^{\mu\nu\alpha}_{abc}
\right\} \omega_\nu{}^{bd} \omega_\alpha{}^{cd}
\end{eqnarray}
This equation is a non-linear one. Moreover, if one consider next to
the linear approximations then one obtains even more non-linear
terms. So it seems hardly possible to get general solution of this
equation, but nothing prevent us from solving it iteratively, order by
order. Here we restrict ourselves by the linear approximation as in
the previous case. In the lowest order approximation we get:
\begin{equation}
h_\mu^{(1)a} = \frac{1}{d-2} \hat{R}_\mu{}^a, \qquad \hat{R}_\mu{}^a =
 R_\mu{}^a - \frac{1}{2(d-1)} e_\mu{}^a R
\end{equation}
and in this notations the structure of quadratic second derivative
Lagrangian looks very similar to the $d=3$ case:
\begin{equation}
{\cal L}_0 = - \frac{1}{2(d-2)} \left\{ \phantom{|}^{\mu\nu}_{ab}
\right\} \hat{R}_\mu{}^a \hat{R}_\nu{}^b + \frac{\kappa}{2}
\left\{ \phantom{|}^{\mu\nu}_{ab} \right\} \omega_\mu{}^{ac}
\omega_\nu{}^{bc}
\end{equation}

The formulas in the next approximation could be greatly simplified
if we introduce traceless conformal Weyl tensor:
\begin{equation}
C_{\mu\nu}{}^{ab} = R_{\mu\nu}{}^{ab} - \frac{1}{d-2} e_{[\mu}{}^{[a}
 R_{\nu]}{}^{b]} + \frac{1}{(d-1)(d-2)} e_\mu{}^{[a} e_\nu{}^{b]} R
\end{equation}
in this, the following useful relation holds:
\begin{equation}
R_{\mu\nu}{}^{ab} = C_{\mu\nu}{}^{ab} + \frac{1}{d-2} e_{[\mu}{}^{[a}
\hat{R}_{\nu]}{}^{b]}
\end{equation}

As in the $d=3$ case it is possible to consider a "flat" limit with
when $k \rightarrow 0$ and $\kappa \rightarrow 0$ keeping $K =
\frac{k}{\sqrt{\kappa}} $ fixed. In this limit a solution of $h$
equation in the next order gives:
\begin{equation}
h_\mu^{(2)a} = - \frac{K}{(d-2)^3} [(d-2) C_{\mu\nu}{}^{ab}
\hat{R}_b{}^\nu -  \hat{R}_\mu{}^\nu \hat{R}_\nu{}^a + 
 \hat{R}_\mu{}^a \hat{R} + 
\frac{1}{2(d-1)} e_\mu{}^a [ (\hat{R} \hat{R}) - \hat{R}^2 ]]
\end{equation}
Then putting this expression back to the initial first order
Lagrangian and keeping only cubic terms we obtain the following
three derivatives Lagrangian:
\begin{equation}
{\cal L}_1 = - \frac{K}{2(d-2)^2} [
 \left\{ \phantom{|}^{\mu\nu\alpha\beta}_{abcd} \right\}
 C_{\mu\nu}{}^{ab} \hat{R}_\alpha{}^c \hat{R}_\beta{}^d +
\frac{d-4}{3(d-2)} \left\{ \phantom{|}^{\mu\nu\alpha}_{abc} \right\}
\hat{R}_\mu{}^a \hat{R}_\nu{}^b \hat{R}_\alpha{}^c ]
\end{equation}
Again this particular structure of the Lagrangian is fixed not only
by the invariance under the usual gauge transformations $\delta
\omega_\mu{}^{ab} = \partial_\mu \eta^{ab}$, but also by the
invariance under the local $\xi$ shifts with the linear terms being:
\begin{equation}
\delta_1 \omega_\mu{}^{ab} = K [ \xi^\nu C_{\nu\mu}{}^{ab} + 
\frac{1}{d-2} ( \hat{R}_\mu{}^b \xi^a - \hat{R}_\mu{}^a \xi^b) ]
\end{equation}
The following identities turn out to be useful:
\begin{eqnarray}
D_a C_{\mu\nu}{}^{ab} &=& \frac{1}{d-2} (D_\mu \hat{R}_\nu{}^b - 
D_\nu \hat{R}_\mu{}^b) \nonumber \\
D_a \hat{R}_\mu{}^a &=& D_\mu \hat{R}
\end{eqnarray}
Note, that the general structure of the Lagrangian obtained is in
agreement with the $d=3$ case. Indeed, in $d=3$ conformal Weyl tensor
is identically zero, so the first term is absent. It is interesting
to note that the $d=4$ case is also special, because in this and only
this case the second term is absent.

\section*{Conclusion}

In this paper we have shown that there indeed exists a dual
formulation of gravity in terms of Lorentz connection
$\omega_\mu{}^{ab}$ field. Such formulation turns out to be highly
non-linear higher derivatives theory, so it is not an easy task (if
at all possible) to give compact formulation at full non-linear
level. However it is possible to construct such theory iteratively,
order by order in fields as we have done in the linear approximation
here. Also we have shown that the so called exotic parity-violating
interactions for massless spin-2 particles could be considered just
as such dual formulation of usual gravitational interactions.

\end{document}